\documentclass[dvipsnames,aps,prx,nobibnotes,twocolumn,superscriptaddress,nolongbibliography]{revtex4-2}[prl]
\usepackage[english]{babel}
\usepackage[utf8]{inputenc}
\usepackage{graphicx}
\usepackage[unicode=true,pdfusetitle,
 bookmarks=true,bookmarksnumbered=false,bookmarksopen=false,
 breaklinks=false,pdfborder={0 0 0},pdfborderstyle={},backref=false,colorlinks=true]
 {hyperref}
\hypersetup{
 citecolor=blue,
 urlcolor=blue,
 linkcolor=blue
}
\usepackage{pdflscape}
\usepackage{latexsym,amsmath,verbatim}
\usepackage{amssymb} 
\usepackage{bm}
\usepackage{amsfonts}
\usepackage{amsmath}
\usepackage{colortbl}
\usepackage{array}
\usepackage{stackrel}
\usepackage{bm}
\usepackage{nicefrac}
\usepackage{rotating}
\usepackage{float}
\usepackage[final]{pdfpages}
\usepackage{lipsum}  
\usepackage{braket}
\usepackage{dsfont}
\usepackage[normalem]{ulem}
\usepackage{xcolor}
\usepackage{nameref}
\usepackage{cleveref}
\usepackage{pdfpages}

\usepackage{multibib}
\newcites{sup}{Supplementary References}

\DeclareMathOperator{\Tr}{Tr}

\makeatletter
\AtBeginDocument{\let\LS@rot\@undefined}
\makeatother

\crefname{figure}{Fig.}{Figs.}
\crefname{equation}{Eq.}{Eqs.}

\begin{document}

\title{Persistent currents in signed directed networks}

\author{Davide Cipollini}
\email[Corresponding author: ]{d.cipollini@inrim.it}
\affiliation{Advanced Materials Metrology and Life Sciences Division, Istituto Nazionale di Ricerca Metrologica, Turin, Italy}

\author{Guido Caldarelli}
\email[Corresponding author: ]{guido.caldarelli@cnr.it}
\affiliation{
 CNR-ISC Institute of Complex Systems, via dei Taurini 19, 00185 Rome, Italy
}%
\affiliation{
 DSMN and ECLT Ca'Foscari University of Venice, via Torino 155, 30172 Venezia Mestre, Italy
}%
\affiliation{%
LIMS London Institute for Mathematical Sciences, Royal Institution, Albermarle st. 21 W1S 4BS London, UK
}%




\begin{abstract}
Network theory can be fruitfully used to describe quantum coherence in physical systems. To that purpose we introduce persistent currents in signed directed networks by interpreting the signed magnetic Laplacian as an effective Hamiltonian and the associated edge phases as a discrete gauge field. In a canonical ensemble, persistent currents arise as thermodynamic responses to variations of gauge-invariant fluxes. We show that these fluxes are naturally defined on the cycle space of the network, and that the resulting currents are constrained to the divergence-free subspace and decompose onto independent cycles. This formulation provides a direct generalization of persistent currents from rings and lattices to arbitrary topologies. Detection of persistent currents provides a signature of the quantum phase coherence supported by the network, and a direct signature of the geometry of its cycle space. Such a mapping, not only allows a practical way to deal with quantum coherence for a variety of situations in the field of quantum technologies, but it also allows a physical interpretation of the importance of the Laplacian operator in graph theory, linking its role to the one of Hamiltonian ({\em i.e.} a tight-binding one) in physical systems. To test the power of the method, we construct a signed directed network that reproduces the Hofstadter butterfly spectrum.
\end{abstract}

\maketitle




Signed and directed networks naturally encode asymmetric and antagonistic interactions across a wide range of systems, from social systems~\cite{heider_attitudes_1946,cartwright_structural_1956} to neural networks~\cite{hopfield_neural_1882,hopfield_physics_1994}. 

While spectral and dynamical approaches based on diffusion kernels have been developed for undirected networks~\cite{Newman2018,Masuda_2017,de_domenico_spectral_2016,villegas_laplacian_2023,cipollini_tree_2025}, magnetic Laplacians~\cite{shubin_discrete_1994} have been used to encode edge orientation through phases and to reveal loop-related structures such as flux communities~\cite{Fanuel2017}. 
More recently, signed magnetic Laplacians have been introduced to incorporate antagonistic interactions and the weighted directional imbalance between opposite edges in the context of machine learning~\cite{singh2023signed}.

However, a thermodynamic framework for loop-based transport and interference in signed directed networks is still lacking. 
The framework we propose here provides a physical interpretation of the graph Laplacian as an effective Hamiltonian, in analogy with the one of physical quantum systems.
\\

Gauge fields provide a natural language to describe orientation and frustration across scales. 
Associating phases to edges, one can encode directed and signed interactions as a discrete gauge field on the network.

This construction generalize the notion of magnetic phases in condensed matter systems~\cite{shubin_discrete_1994,Peierls1933}, where the accumulation of phase along closed paths gives rise to observable effects such as persistent currents~\cite{Bergmann1984} and flux quantization~\cite{London1950,Onsager1961}. 
Interestingly, persistent currents emerge in systems that showcase a macroscopic coherence, such as superconductors~\cite{tinkhamintroduction} and superfluids~\cite{London1950}, but they can also be observed in micrometer-scale mesoscopic metallic rings~\cite{imry1998introduction} and in synthetic magnetism with application to quantum technologies~\cite{Amico2022}.


In this work, we show that persistent currents can be defined on arbitrary signed directed networks as thermodynamic responses to gauge-invariant fluxes. 
Such fluxes arise as holonomies of the edge phases, {\em i.e.,} a measure of how much a geometric or gauge structure ``twists'' when you move around a loop. They are naturally associated with the cycle space of the network, which emerges as the fundamental domain governing transport and interference, consistent with the absence of sources and sinks at the nodes.

This perspective is directly relevant to physical platforms with controllable gauge phases.
In fact, while it provides a multiscale signature of phase coherence and loop interference in signed directed networks, it offers a natural language for phase-coherent physical platforms admitting an effective network description, including superconducting circuits such as Josephson-junction arrays and SQUID networks, as well as synthetic quantum matter platforms~\cite{Dalibard2011,Goldman2014,Amico2022} such as ultracold atoms~\cite{DelPace2022} with application to atomtronic circuits. In particular, recent developments have focused on engineering exotic geometries in synthetic quantum matter to simulate phenomena ranging from topology to analog gravitational and cosmological effects~\cite{Grass2025}. 

Therefore, extending persistent currents to such complex geometries through networks provides a novel perspective readily applicable to the rapidly expanding field of quantum technologies and high precision sensing.

We present in the following some analytical results of this thermodynamic construction.

\begin{figure}[hbtp]
    \centering
    \includegraphics[width=\columnwidth]{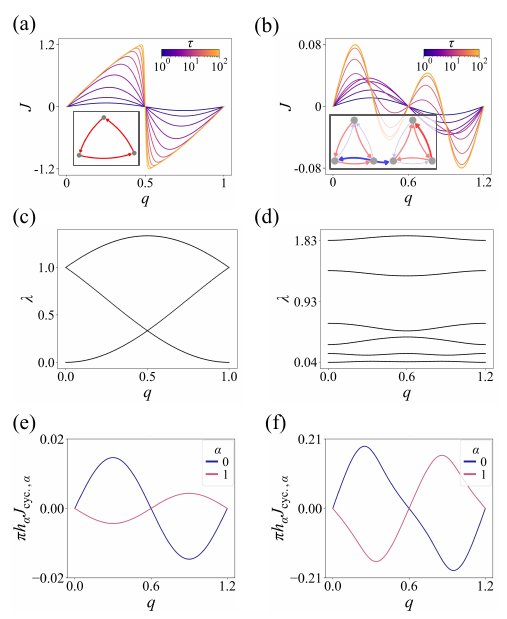}
    \caption{Persistent currents in looped networks. 
(a) Current as a function of $q$ for a single-cycle network (inset), showing periodic behavior. 
(b) Current for a network with two independent cycles (inset).
(c,d) Corresponding spectra of the signed magnetic Laplacian.
Panels (e) and (f) show the contributions of the two cycles to the global current at $\tau=1$ and $60$ respectively, highlighting the multiscale nature. The insets in (a) and (b) show the networks: weights are color coded in blue (negative) and red (positive). Transparency codifies for amplitude.}
    \label{fig:persistent_currents_loops}
\end{figure}

\paragraph*{Thermodynamic currents on networks\textemdash}
We consider a signed directed network described by the signed magnetic Laplacian $L^{(M)} = D^{(M)} - A^{(M)}$, where the magnetic adjacency matrix is defined as
\begin{equation}
A^{(M)}_{ij} = A^{\mathrm{sym}}_{ij}\, e^{i\phi_{ij}},
\qquad
\phi_{ij} = \pi q\, \Delta_{ij},
\end{equation}
with $A^{\mathrm{sym}}_{ij} = \tfrac{1}{2}(A_{ij}+A_{ji})$, $\Delta_{ij}=A_{ij} - A_{ji}$ and degree matrix $D^{(M)}_{ij}=\delta_{ij}\sum_j |A^{\mathrm{sym}}_{ij}|$~\cite{singh2023signed}. 
This construction encodes edge orientation and sign as a discrete gauge field on the network, generalizing the Peierls substitution to arbitrary graphs~\cite{shubin_discrete_1994,lieb_fluxes_2004}. 
Up to a diagonal term, $L^{(M)}$ is equivalent to a tight-binding Hamiltonian in the presence of a magnetic field, where particles acquire a phase along each edge. Thus, its spectrum differs from that of the corresponding Hamiltonian only by a constant shift (refer for more details to the Supplemental Material (SM)).
Moreover, it reduces to the signed Laplacian~\cite{Kunegis2010,iannelli_topological_2025} in the absence of asymmetry, and to the standard Laplacian when the network is unsigned. 

We adopt a density matrix formulation for networks~\cite{de_domenico_spectral_2016}, which provides a natural framework to define thermodynamic potentials~\cite{ghavasieh_diversity_2024}. The Gibbs density matrix reads
\begin{equation}
\rho = \frac{e^{-\tau L^{(M)}}}{Z}, 
\qquad 
Z = \Tr\, e^{-\tau L^{(M)}},
\end{equation}
where $\tau$ plays the role of an inverse temperature or diffusion time~\cite{villegas_laplacian_2023}, and the corresponding free energy is
\begin{equation}
F = -\frac{1}{\tau} \ln Z.
\end{equation}
The positive semidefiniteness of $L^{(M)}$~\cite{singh2023signed} guarantees that the Gibbs ensemble defines a probability measure over its set of eigenstates.

In mesoscopic rings, persistent currents arise as derivatives of the free energy with respect to the magnetic flux threading the system~\cite{Bttiker1983}. Thus, they can be understood as thermodynamic responses to gauge-invariant fluxes. 
This motivates a generalization through signed directed networks, where fluxes are encoded through edge phases and parametrized by the quantity $q$.

We define the global persistent current as
\begin{equation}
\label{eq:global_current}
J(q) := -\frac{1}{\tau}\frac{d}{d q} \ln Z(q).
\end{equation}

Notably, the spectrum of $L^{(M)}$ is invariant under gauge transformations of the form $\phi \;\mapsto\; \phi + B^\top \chi$, where $B\in\mathbb{R}^{N\times E}$ is the node-edge incidence matrix and $\chi = (\chi_1,\dots,\chi_N)$ is a scalar function on the $N$ nodes (see SM). Therefore, the free energy depends only on gauge-invariant combinations of phases. 

These gauge-invariant combinations are naturally expressed in terms of cycle fluxes. 
Introducing the cycle-edge incidence matrix $C \in \mathbb{R}^{\beta_1 \times E}$, where $E$ is the number of undirected edges of the support graph~\footnote{Here, the support graph is the undirected graph obtained by ignoring edge orientation, so that each undirected edge $e=(i,j)$ represents the pair of directed edges $(i\to j)$ and $(j\to i)$.} and $\beta_1$ is the number of independent cycles, we define
\begin{equation}
\Psi = C \phi,
\end{equation}
which assigns a total phase (holonomy) to each cycle. 
Note that the rows of $C$ form a basis of the cycle space $\ker (B) \subset \mathbb{R}^E$.

The free energy can therefore be regarded as a function of $\Psi$, and the corresponding currents are the conjugate variables to the cycle-fluxes
\begin{equation}
\label{eq:J_cycle_conj_var}
J_{\mathrm{cycle}} = -\frac{1}{\tau} \nabla_{\Psi} \ln Z(\Psi).
\end{equation}

We may now  decompose the global current in \Cref{eq:global_current} highlighting each cycle' contribution: $J(q)=\frac{dF}{dq}=
\sum^{\beta_1}_{\alpha} \frac{\partial F}{\partial{\Psi_\alpha}} \frac{d\Psi_\alpha}{dq}=\pi h^\top J_{\mathrm{cycle}}$, where $h_\alpha =\ (C\Delta)_\alpha$. Under this decomposition, the global current can be regarded as the directional derivative of the free energy along the $h$ direction in cycle-holonomy space, or equivalently the sum of the cycle currents $J_{\mathrm{cycle}}$ weighted by $h$.

To resolve the structure of the currents on the network edges, we may introduce edge currents as responses to variations of the edge phases,
\begin{equation}
j_e := -\frac{1}{\tau} \frac{\partial \ln Z}{\partial \phi_e}.
\end{equation}

Using the chain rule \footnote{Since the free energy $F$ depends on the edge-phases $\phi_e$ only through the gauge-invariant cycle holonomies $\Psi=C\phi$, one has $
\partial_{\phi_e}F
=
\sum_\alpha
\frac{\partial F}{\partial \Psi_\alpha}
\frac{\partial \Psi_\alpha}{\partial \phi_e}
=
\sum_\alpha
\frac{\partial F}{\partial \Psi_\alpha}
C_{\alpha e}$.
Therefore, $j_e= C^\top J_{\mathrm{cycle}}$
}, the edge currents can be expressed as
\begin{equation}
j = C^\top J_{\mathrm{cycle}}.
\end{equation}

Since $B C^\top = 0$, it follows that $B j = 0$, implying current conservation at nodes. 
Persistent currents are therefore divergence-free equilibrium currents and naturally decompose onto the cycle basis, identifying the cycle space $\ker (B)$ as the topological domain governing transport and interference in this setting.

\paragraph*{Loop currents and interference\textemdash}
We illustrate the emergence of persistent currents on two simple network topologies containing closed loops in ~\Cref{fig:persistent_currents_loops}. The Figure showcases the responses to the global phase parameter $q$ for networks with one and two independent cycles.
In panel (a), while at low temperature the cuspid in the lower energy level directly produces the sawtooth current, at higher temperatures, crossing of the levels leads to the gradual evolution of the current from sawtooth to the sinusoidal.

The global current exhibits periodicity in $q$, reflecting the fact that the spectrum depends only on the phase accumulated along the cycles, i.e., the holonomies. 
For a single cycle $\alpha$, the accumulated phase is $\Psi_\alpha = \pi q h_\alpha$, leading to a fundamental period $q_\alpha = 2/|h_\alpha|$. For a multi-cycle network, to have periodicity, all accumulated phases must return to multiples of $2\pi$.
When the edge profile $\Delta$ is unitary, $h_\alpha$ reduces to the cycle length, and the periodicity scales accordingly. 
This generalizes the periodic dependence of persistent currents on magnetic flux from mesoscopic rings to arbitrary network topologies.


In networks with multiple loops, the current inherits the rich structure arising from the multiscale interference between cycle currents. 

The thermodynamic parameter $\tau$ controls the relative weight of the contributing modes, thereby determining whether the current response is approximately local or exhibits long-range correlations on edges. In fact, currents are not independent on edges, but they are constrained by the topology of the network and organized along cycles. 

For $\tau \to 0$, one has the maximally mixed state $\rho = I/N + \mathcal{O}(\tau)$, so that correlations between different edge perturbations are suppressed and the response becomes approximately local in edge space. 
For $\tau \to \infty$, the Gibbs state is dominated by the lowest mode, $\ln Z \sim -\tau \lambda_0(q)$, so that the current is controlled by the sensitivity of the ground state to the gauge field, $J(q)\sim \partial_q \lambda_0(q)$. 

Interference between cycles sharing edges is a topological effect that encodes the geometry of the cycle space.
This structure is analogous to a circuit of coupled inductors, where currents interact through mutual inductances.

This can be made explicit by expanding the free energy in the small-flux regime,
\begin{equation}
\label{eq:linear_exp_free_energy}
F(\Psi) = F(0) + \frac{1}{2} \Psi^\top \mathcal{K}(\tau)\, \Psi + \mathcal{O}(\Psi^4).
\end{equation}
where $\mathcal{K}(\tau) =\nabla^2_\Psi F _{|\Psi=0}$ is the response kernel with respect to the coordinates $\Psi_\alpha$ of the independent
cycle holonomies associated with the chosen cycle basis. 
It can be interpreted as an effective phase stiffness tensor, as it quantifies the energetic cost due to variations of the gauge-invariant fluxes.

Deriving by the cycle coordinates (\Cref{eq:J_cycle_conj_var}), we obtain the linear relation between currents and fluxes $J_{\mathrm{cycle}} = \mathcal{K}(\tau)\, \Psi$. 
This linear relation, jointly with \Cref{eq:linear_exp_free_energy}, indicates that $\mathcal{K}(\tau)^{-1}$ plays the role of an effective inductance matrix, making the system formally analogous to a circuit of coupled inductors. This points to a formal duality between cycle stiffness and inductive coupling.

At the edge level, we introduce the phase-response kernel $\Gamma=\nabla^2_\phi F$, whose elements $\Gamma_{ee'}$ measure the response of the current on edge $e$ to a phase perturbation on edge $e'$. Projecting this kernel onto the gauge-invariant cycle space gives (see SM)
\begin{equation}
\label{eq:cycle-cycle_response}
\mathcal{K}(\tau)=M^{-1} C\,\Gamma(\tau)\,C^\top M^{-1}.
\end{equation}
where $M = C C^\top$ is the Gram matrix of the chosen cycle basis, whose off-diagonal entries quantify the signed and purely geometric overlap between cycles in edge space.

In fact, although the cycles form a basis of $\ker (B)$ and are therefore linearly independent, they are in general not orthogonal in edge space, as different cycles may share edges. 
In this case, $M$ becomes non-diagonal, inducing a non-Euclidean metric that controls how currents associated with different cycles interfere on edges.

For $\tau \to 0$, $\Gamma$ becomes approximately diagonal, as the nonlocal off-diagonal components of the edge-phase response kernel are suppressed (see SM). 
If, in addition, the local edge responses are sufficiently uniform after projection onto the gauge-relevant cycle subspace, i.e., $C\Gamma(\tau)C^\top \simeq \gamma(\tau)CC^\top$, than one obtains that the effective inductance matrix is proportional to the geometrical cycle Gram matrix, $\mathcal K(\tau)^{-1}\simeq \gamma(\tau)^{-1}M$. 

At finite $\tau$, however, $\Gamma(\tau)$ is generally neither local nor uniform. In this regime, the coupling between cycle responses is no longer controlled only by the pure geometry of the cycle space through $M$, but also by the structure of the edge response kernel $\Gamma$. Nonlocal components of $\Gamma$ couple perturbations applied on different edges and therefore generate additional cycle-cycle couplings through~\Cref{eq:cycle-cycle_response}.
These contributions are spectral in origin and depend on how the eigenstates of $L^{(M)}$ overlap with the edge perturbation operators.

This aspect is highlighted in \Cref{fig:persistent_currents_loops}(b), where the two cycles do not share edges and, consequently, the cycle-geometric Gram matrix $M$ is diagonal. Therefore, the coupling between cycle responses cannot originate from direct cycle overlap, but from the nonlocal spectral response kernel $\Gamma(\tau)$, which couples edge perturbations through eigenstates of the magnetic Laplacian that extend over the whole network. As a result, perturbations on different cycles become effectively coupled, leading to a smooth dependence of the free energy on the phase parameter. 
The resulting smooth current persists even in the limit $\tau \rightarrow \infty$, indicating that it originates from eigenstate delocalization, which can lift degeneracies and remove level crossings in the energy spectrum in \Cref{fig:persistent_currents_loops}(d).

\begin{figure}[]
    \centering
    \includegraphics[width=\linewidth]{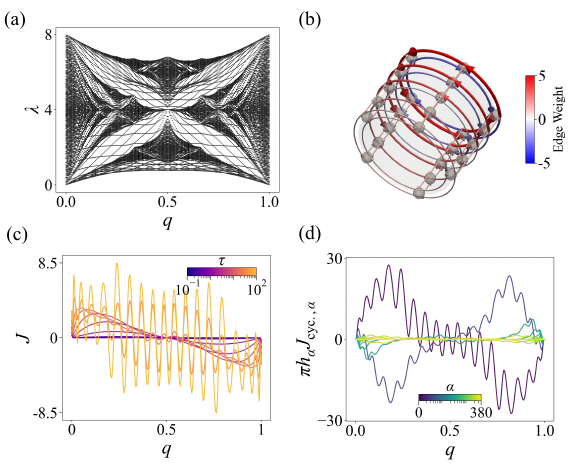}
    \caption{Hofstadter network. 
(a) Energy spectrum computed on a network with $N=400$ nodes as a function of $q$, showing the butterfly-like fractal structure. 
(b) Representative signed directed network with spatially structured asymmetric weights, shown for a smaller system with $N=25$ nodes for visualization purposes. 
(c) Global persistent current computed on the full $N=400$ network. 
(d) Cycle-resolved currents at $\tau=10^2$ computed on the same $N=400$ network.}
    \label{fig:hoftsadter_network}
\end{figure}

\paragraph*{Hofstadter network\textemdash}
As an application of the proposed framework, we construct a signed directed network that reproduces a Hofstadter-like butterfly spectrum~\cite{Hofstadter}. 
In this construction, the effect of a magnetic field is encoded through a spatially structured pattern of asymmetric edge weights described in the SM, and corresponding to a discrete Landau gauge field defined on the network.

Figure~\ref{fig:hoftsadter_network} shows the spectrum as a function of $q$. 
Its fractal structure arises from interference between currents on correlated cycles and can be described purely in terms of cycle fluxes, without relying on plaquette-based geometric notions.

This demonstrates that nontrivial magnetic phenomena, typically associated with regular lattices and continuum gauge fields, can be captured in purely topological terms on arbitrary networks. The phase structure of the signed magnetic Laplacian eigenvectors in the Hofstadter network is reported in the SM, showing how the eigenstate phases reorganize as the control parameter $q$ is varied. Moreover, our work exposes the tight relation between Hamiltonians and network Laplacians (see for further discussion the SM).

\paragraph*{Conclusion\textemdash}
We have introduced a framework to define persistent currents on signed directed networks as thermodynamic responses to gauge-invariant cycle fluxes. 
This formulation identifies the cycle space as the natural domain for equilibrium transport and interference in the absence of sources and sinks, extending the notion of persistent currents from rings and lattices to arbitrary network topologies.

Our results provide a bridge between spectral network theory, gauge theory, and phase-coherent transport in physical systems. 
They open new directions for the multiscale characterization of coherent response and frustration in asymmetric and antagonistic complex systems. 
More broadly, the cycle-flux formulation may be relevant for phase-coherent platforms with controllable gauge phases, including Josephson networks, artificial quantum matter, analog quantum simulators, and ultracold-atom systems relevant to atomtronics and high precision sensing~\cite{Dalibard2011,Goldman2014,Amico2022}.

\paragraph*{Outlook\textemdash}
Extensions beyond the present equilibrium setting should address driven or open phase-coherent networks by including externally imposed currents, voltage biases, dissipation, or time-dependent gauge fields. 
This would provide a novel graph-based perspective of quantum-information processing via superconducting or mesoscopic devices operating out of equilibrium.

More speculatively, a non-equilibrium extension could provide a route to interpret learning in artificial neural networks in analogy to the driven response of an adaptive phase-coherent networked system. 
In such a picture, data would enter as external drives, boundary conditions, or imposed gauge phases, while the system makes inference through the resulting pattern of currents and cycle fluxes. 
Learning would then correspond to adapting the network couplings or gauge structure so as to shape this response, suggesting a possible bridge towards the thermodynamic interpretability of deep learning. 
Finally, beyond the physical metaphor, the phase-coherent system may no longer only be meant as an analogy for computation, but become an actual physical system whose intrinsic network dynamics performs information processing~\cite{Markovic2020}. 
Gauge-controlled coherent responses and loop interference could then serve as phase-sensitive degrees of freedom for physics-grounded information processing in magnetic, mesoscopic, or superconducting materials.

\section*{Acknowledgments}
DC and GC are grateful to Andrea Puglisi for insightful discussions during the early stages of this work, and to Gianluca Milano for useful comments on a later version of the manuscript.
GC acknowledges Cognia Visitors scheme in LIMS. 

\bibliography{references_main}







\end{document}